 \documentclass[twocolumn,epjc3]{svjour3}  
\smartqed  
\RequirePackage{graphicx}

\usepackage{graphicx}
\usepackage{dcolumn}
\usepackage{bm}
\usepackage{amssymb}
\usepackage{amsmath}
\usepackage{hyperref}
\usepackage{multirow}
\usepackage{cancel}

\usepackage[dvipsnames]{xcolor}

\usepackage{pifont}

\hypersetup{
    colorlinks,
    citecolor=blue,
    filecolor=black,
    linkcolor=blue,
    urlcolor=blue
}
\journalname{}
\date{}
\begin{document}

\title{Reinterpreting Neutrino Oscillations}
\author{Marc Bergevin\thanksref{e1,addr1}}
\thankstext{e1}{e-mail: bergevin1@llnl.gov}
\institute{Lawrence Livermore National Laboratory\label{addr1}}

\maketitle
\begin{abstract}

This letter proposes an alternative quantum mechanical picture for the observed phenomena of neutrino oscillations. It is assumed in the following that neutrinos interact via diabatic (or localised) interactions with a new particle field, which changes their flavor. 
Furthermore, it is assumed that each neutrino flavor state can only have a single associated mass thereby making them fundamental particles of nature. The effective masses associated with matter interactions replace the concept of neutrino mixing angles. Preliminary evidence that left-handed neutrinos and right-handed antineutrinos oscillate differently is presented, implying charge-parity violation.
Given the apparent anomalous observations of some neutrino oscillation experiments, which have led to speculations about the existence of a fourth (sterile) neutrino, it is worth examining the oscillation behavior predicted by alternative mechanisms to determine if they more naturally explain the available data.

\end{abstract}

\maketitle

Contrary to charged leptons, neutral leptons  are thought to be produced as a quantum superposition of at least three mass states \cite{FRITSCH197672,PhysRevD.98.030001}. Over the last 20 years, the community has made great progress in measuring the oscillation properties of the Pontecorvo Nakagawa Maki Sakata (PNMS) system  \cite{Schwetz_2008,review}. Flavor oscillations are assumed to occur due to an energy difference in the mass states leading to oscillations in the interaction states as a function of time. However, experimental tensions have arisen in multiple neutrino sectors in the last 20 years. For example, the SAGE/GALLEX \cite{PhysRevC.80.015807,Gavrin_2011} results are outside of the predicted oscillation expectation and are in conflict with the Borexino results at similar energies \cite{PhysRevD.89.112007}. More recently, new reactor antineutrino spectra and a re-evaluation of the neutron lifetime highlighted that short-baseline reactor-antineutrino results were systematically lower than expected---by 6\%---implying potential oscillations at short baselines \cite{PhysRevD.83.073006}. Recent re-evaluation of the spectral conversion of electron to electron antineutrinos have resulted in an upward shift of 3\% ($\Phi_{corr}$), partly alleviating the problem  \cite{PhysRevD.98.030001}.
Spectral features at 5 MeV  \cite{PhysRevLett.114.012502,PhysRevD.99.055045} are present in the reactor data, though these same oscillation features cannot be confirmed by recent searches \cite{PhysRevLett.121.251802} and this anomaly is still an open question. Finally, the LSND and MiniBoone experiments \cite{ATHANASSOPOULOS1997149,PhysRevLett.121.221801} observed an excess of electron neutrinos at lower energies, this excess is not in agreement with the accepted model of oscillations \cite{Goodman:2019gin,PhysRevD.85.092008}. While errors in the flux models, cross-section models, interaction effects, or other effects are possible, this letter investigates whether a different neutrino oscillation interpretation might resolve these tensions. 

Non-standard interaction (NSI) models have been proposed for lepton flavor violation effects that should be investigated in conjunction with the PNMS model. A model proposed by Ge and Murayama \cite{ge2019apparent} predicts that a lepton violating process can occur through interactions with dark matter, thus leading to second order oscillation effects observable by the next generation of neutrino experiments. De Gouvea {\it et al} \cite{PhysRevD.100.075033} have reviewed non-standard interactions that could result in charged lepton flavor violation. In this letter we explore the idea that 
flavor oscillations occur as a consequence of 
perturbative interactions with vector bosons, rather than 
neutrino state superposition as assumed in the PNMS model. 
Furthermore, as neutrinos do not interact with regular matter strongly it is assumed that NSI are dominant and that these vector bosons are spread uniformly in space. 

While the Higgs mechanism provides a way for particles to acquire mass, it is not well understood under what conditions the mechanism applies. Is the perturbation {\it adiabatic} or {\it diabatic} in nature? In other words, is mass the by-product of a slow constant perturbation due to an intrinsic property or is it instead a fast localized external perturbation leading to a change of quantum state? We explore here the view that the mechanism can either be {\it adiabatic} or {\it diabatic} for standard mass interaction, but is assumed {\it diabatic} by nature for flavor changes. This assumption is made to simplify the system so that it can be studied with simple dynamical system methods. 

Noble and Jentschura \cite{PhysRevA.92.012101} investigated the ultrarelativistic limit for a perturbative scenario where successive matter fields are acting on the kinetic terms instead of vice-versa. {\color{black} Part of their treatment is adapted here.} The Hamiltonian for a free {\color{black} Weyl spinor} in the presence of a matter field $m$ in the Weyl basis is:
\begin{equation}
\mathbf{H}_{FD} =\begin{cases}  \left(\begin{matrix}
    -\Vec{\sigma}\cdot\Vec{p}\;\;&\;\;m\\
    m&\Vec{\sigma}\cdot\Vec{p}
\end{matrix}\right)\text{  continuous matter field}\\
\left(\begin{matrix}
    -\Vec{\sigma}\cdot\Vec{p}\;\;&\;\;m^{LR}\\
    m^{RL}&\Vec{\sigma}\cdot\Vec{p}
\end{matrix}\right)\text{  annihilation-creation}
\end{cases}
\label{Left-Hand-Oscillation}
\end{equation}
{\color{black} where the first case is the standard introduction of an intrinsic Dirac mass} and {\color{black} the second case introduces an effective mass due to spin flips done via successive application of the annihilation-creation operators. 
While both cases have different underlining physical assumptions, they have the same mathematical solution. The effective mass case is} interpreted as a massless spin 1/2 particle of energy $p$ that undergoes a handedness spin-flip in the presence of a field potential term ($m=m^{LR}=m^{RL}$) where $LR$ and $RL$ are introduced to denote transition energy from left to right states and vice-versa. {\color{black} The system may, in reality, be a mix of intrinsic and external effects. 

The Hamiltonian of this system is,
\begin{eqnarray}
\mathcal{H}_{FD} &=&  \left(\begin{matrix}
    -\frac{\Vec{\sigma}\cdot\Vec{p}}{|\Vec{p}|}\sqrt{\vec{p}^2+m^2}&0\\
    0 &\frac{\Vec{\sigma}\cdot\Vec{p}}{|\Vec{p}|}\sqrt{\vec{p}^2+m^2}
\end{matrix}\right),
\end{eqnarray}
{\color{black} with energy state solutions,
\begin{eqnarray}\mathcal{H}_{FD}\left(
\begin{matrix}
\Psi_1\\
\Psi_2\\
\Psi_3\\
\Psi_4
\end{matrix}\right) &=&
\left(\begin{matrix}
E_{\nu} & 0 & 0 & 0\\
0 & -E_{\nu} & 0 & 0\\
0 & 0 & -E_{\nu} & 0\\
0 & 0 & 0 & E_{\nu}\\
\end{matrix}\right)\left(
\begin{matrix}
\Psi_1\\
\Psi_2\\
\Psi_3\\
\Psi_4
\end{matrix}\right)\nonumber\\
&=&E_{\nu}\eta\left(\begin{matrix}
\Psi_1\\
\Psi_2\\
\Psi_3\\
\Psi_4
\end{matrix}\right)
\end{eqnarray}
where $E_{\nu}=\sqrt{\Vec{k}^2+m^2}$, where $\Vec{k}$ is the physical momentum. In the ultra-relativistic regime,
\begin{equation}
    E_{\nu}=\sqrt{p^2+m^2}\sim p+\frac{m^2}{2p}
    \label{momentumRelation}
\end{equation}
where $|\Vec{k}|\equiv p$ is defined in order to be consistent with special relativity nomenclature. This is the  Quantum Mechanical scenario of the Foldy-Wouthuysen representation \cite{PhysRev.78.29}, in which the Hamiltonian operator is redefined as,
\begin{eqnarray}
&& H \equiv \beta m + \alpha \cdot p \rightarrow H' \equiv \beta E_{p}.
\end{eqnarray}
The positive energy states $\Psi_1$ and $\Psi_4$ represent respectively the left-handed neutrinos and right-handed antineutrinos, and the negative energy states $\Psi_2$ and $\Psi_3$ represent the left-handed antineutrinos and right-handed neutrino \cite{PhysRevA.92.012101}. 



\section*{Vector boson coupling}
An open question remains, What is the origin of the three distinct flavor masses of each fermion family? Could the coupling of a boson and a fermion form an Anyon, which could provide a solution to this question? While they have no charge, the neutrinos can acquire a magnetic dipole moment, or other effects such as the anapole moment, if the neutrino has mass. Such effects would permit effective gauge photon interactions.

The simplest Lagrangian to consider is the Yang-Mill Lagrangian of the coupling two spin-1/2 fields ($\Psi_1$ and $\Psi_2$) of mass $m$ with a spin-1 coupling (with formalism taken from \cite{Griffiths:1987tj}) has the form,
\begin{eqnarray}
    \mathcal{L} &=& \left[i\hbar c \bar{\Psi}\gamma^\mu\partial_\mu\Psi  -m\bar{\Psi}\Psi\right] - \frac{1}{16\pi}\vec{F}^{\mu\nu} \vec{F}_{\mu\nu} \nonumber\\
    &&-(f_{12}\bar{\Psi}\gamma^\mu\vec{\tau} \Psi)\cdot \vec{A_\mu} 
    \label{LagrangianFlavon}
 \end{eqnarray}
where $A_\mu$ are three massless vector gauge fields,  the three-vector notation denotes the particle index and not spatial coordinates, $\tau$ are the Pauli matrices, and $f_{12}$ is a coupling constant of state 1 to 2\footnote{analogous to the electric charge}. Here state 1 and state 2 are said to have the same mass, but are assumed to have different flavors.  The three gauge particle fields ($x$,$y$,$z$) are,
\begin{equation}
    \vec{F}^{\mu\nu} = \partial^\mu\vec{A}^\nu - \partial^\nu\vec{A}^\mu - \frac{2f_{12}}{\hbar c}\left(\vec{A}^\mu \times \vec{A}^\nu\right),
\end{equation}
with $\Psi$ defined as,
\begin{equation}
    \Psi = \begin{bmatrix}\Psi_1\\ \Psi_2 \end{bmatrix}, \bar{\Psi}=\begin{bmatrix} \bar{\Psi}_1 \hspace{2mm} \bar{\Psi}_2  \end{bmatrix} .
\end{equation}

Here the three possible state interactions are, 
\begin{equation}
    \tau_x \Psi  = \begin{bmatrix}
        \Psi_2\\\Psi_1
    \end{bmatrix},\tau_y \Psi = \begin{bmatrix}
        -i\Psi_2\\i\Psi_1
    \end{bmatrix},\tau_z \Psi = \begin{bmatrix}
        \Psi_1\\-\Psi_2
    \end{bmatrix},
    \label{currents}
\end{equation}
in which an $x$ interaction will  change flavor state 1 to flavor state 2 (F); $y$ will change state 1 to state 2 and provide a switch to the negative energy state (FP); $z$ will provide a switch to the negative energy state for one of the two particles (P).
\begin{figure}[h]
\includegraphics[width=0.48\textwidth]{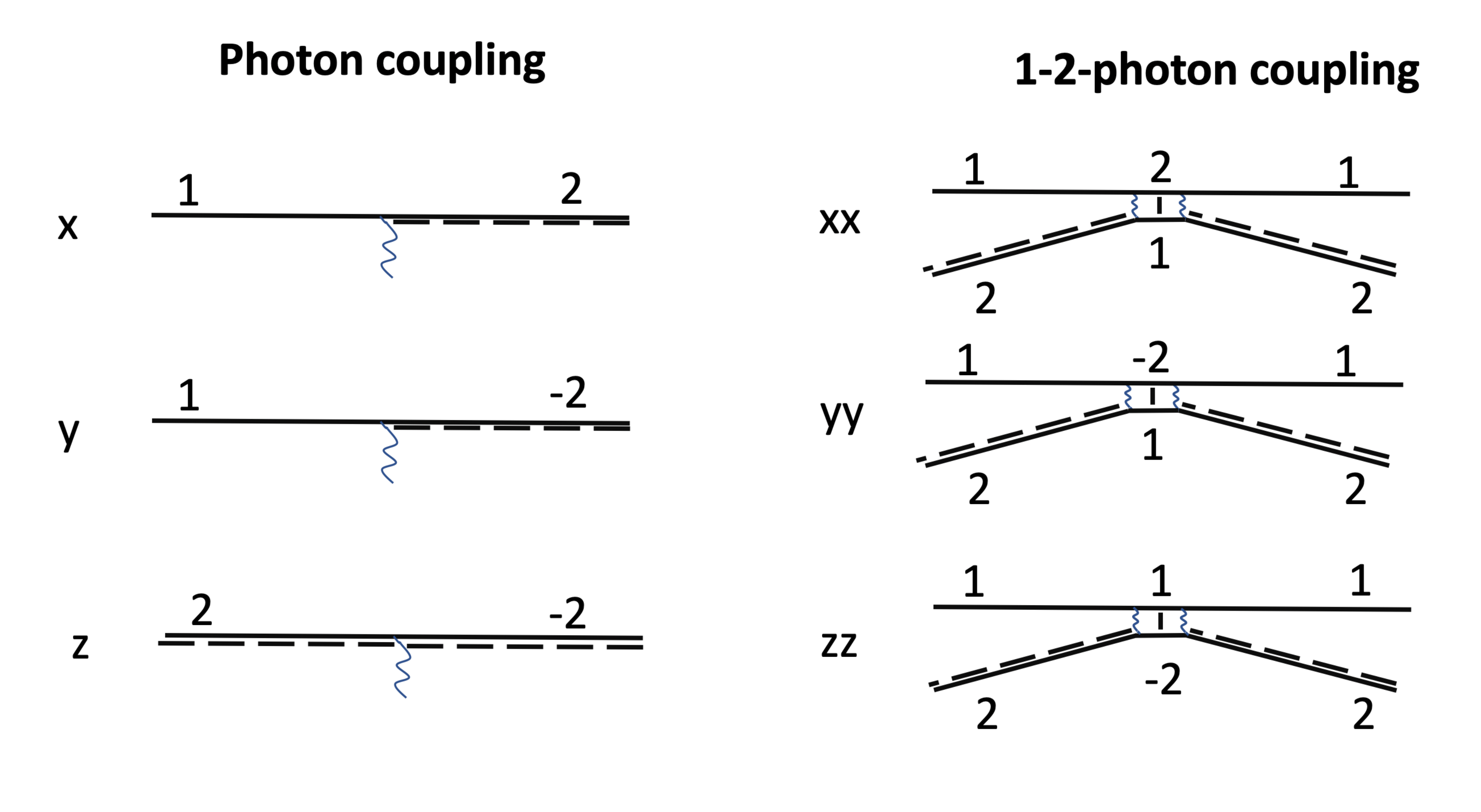} 
\caption{{\color{black} Three possible current interactions (x,y,z) assuming state 2 is the Anyon state of a spin-1/2 (solid line) and a spin-1 field (dashed line). Horizontal lines represent the bound state, and the squiggly lines represent the flavor current exchange. Double current exchange of the same type returns states to their original form, akin to the seesaw mechanism.}} 
\label{bergevinFigure1}
\end{figure} 

 These three interaction currents are shown in Fig.~\ref{bergevinFigure1}. Two successive interactions with the same flavon would reproduce the seesaw mechanism for generic small masses. If the interaction points are localised (diabatic), then one only needs to solve for the energy solution of the non-bound and bound states independently.

\section*{Non-flavor changing interactions}
The previous derivation details how the gauge photons interact with spin-1/2 particles at interaction points. An Anyon formed of the free Weyl spinor and spin-1 particle is assumed to be possible even in the absence of an electric charge. This section concentrates on non-flavor varying current $f_{ij}\tau_z\Psi$.

The classical Hamiltonian for a charged particle in an electric field is, 
\begin{equation}
    H = \sqrt{(\vec{p} - q\vec{A})^2+m ^2}+ q A^0,
    \label{ClassicalHamiltonian}
\end{equation}
which is provided for comparison purposes. The neutrino does not have an electric charge ($q A^0\rightarrow 0$) and this formula does not reflect the nature of mass as assumed in this article. 

The Lagrangian of Eqn~\ref{LagrangianFlavon} can be rewritten as,
\begin{eqnarray}
    \mathcal{L} &=& \left[ \bar{\Psi}\cancel{p}\Psi - \bar{\Psi}f_{ij}\cancel{A}_{k}\cdot\tau^{k} \Psi  -m\bar{\Psi}\Psi\right] - \frac{1}{16\pi}\vec{F}^{\mu\nu} \vec{F}_{\mu\nu}.
    \label{LagrangianFlavon2}
 \end{eqnarray}
For $\tau_z\Psi$ non-flavor changing currents this can be rewritten as,  
\begin{eqnarray}
    \vec{\sigma}\cdot\vec{p}&\rightarrow&\vec{\sigma}\cdot\left(\vec{p}-f_{ij}\vec{A}_{z}\right)\equiv  \vec{\sigma}\cdot\vec{p'}\\
    m&\rightarrow&m + f_{ij} A^0_{z} \equiv m',
\end{eqnarray}
where the three-vector notation now denotes the standard spatial coordinates and ${z}$ the particle index.  Replacing ($\vec{p'}$, $m'$) in Eqn.~\ref{Left-Hand-Oscillation}, the Foldy-Wouthuysen transform has the form,
\begin{eqnarray}
H^{z} &=& \eta \sqrt{(\vec{p}-f_{ij}\vec{A_{z}})^2+(m + f_{ij} A^0_{z})^2}\nonumber\\
&\approx& \eta \left({|\vec{p}|-f_{ij}|\vec{A_{z}}|\cos\theta_{\hat{p}\cdot\hat{A}}+\frac{f_{ij}^2\vec{A_{z}}^2+(m + f_{ij} A^0_{z})^2}{2|\vec{p}|}}\right),\nonumber\\
\end{eqnarray}
where $|\vec{p}|\gg f_{ij}|\vec{A_{z}}|$. Choosing $\vec{A}$ to be perpendicular\footnote{which may be true on average i.e. $<\cos\theta_{\hat{p}\cdot\hat{A}}>=0$ if angle is not quantized.} to $\vec{p}$ ensures that no charge term remains in the Hamiltonian satisfying the necessary condition for neutrinos. Moreover, defining the electric charge as $q\equiv -\cos\theta_{\hat{p}\cdot\hat{A}}$ allows $(q=0)$ when $\vec{A}$ is perpendicular to $\vec{p}$, $(q=-1)$ when $\vec{A}$ is parallel to $\vec{p}$, and $(q=1)$ when $\vec{A}$ is antiparallel to $\vec{p}$. 

In general the total ultrarelativisic Hamiltonian then has the form, 
\begin{equation}
    H^{z} \approx \eta \left( |\vec{p}|  +\frac{m_{z}^2}{2|\vec{p}|}+\dots + q f_{ij} |\vec{A}_{z}| \right),
\end{equation}
which is the updated ultra-relativistic form of Eqn.~\ref{ClassicalHamiltonian}. The neutrino ($q=0$) case the total Hamiltonian has the form, 
\begin{equation}
    H^{z} \approx \eta \left( |\vec{p}|+\frac{m_{z}^2}{2|\vec{p}|}+\dots\right),
\end{equation}
which is consistent with Eqn.\ref{momentumRelation} with an effective mass term of,
\begin{equation}
    m_{z} \equiv \sqrt{ (m + f_{ij} A^0_{z})^2 + |f_{ij}\vec{A_{z}}|^2 }.
\end{equation}
Allowing the extension of a third spin-1/2 state, leads to three possible mass splitting ($f_{12},f_{23},f_{13}$). We consider this the origin of the three-flavor mass of spin-1/2 fermions. 
 }
 \begin{figure*}[htp]
\centering
\includegraphics[width=0.98\textwidth]{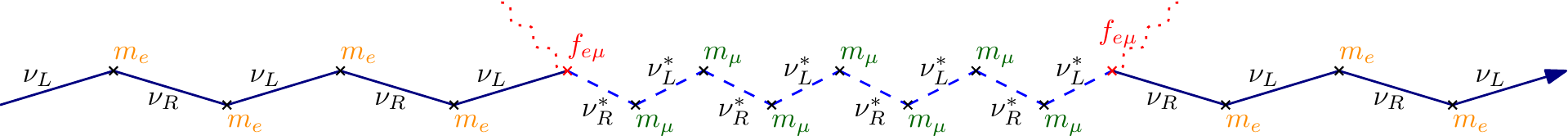} 
\caption{Oscillation picture using diabatic (fast) interactions, denoted by x, with the higgs and flavon boson assumption (diagram adapted from \cite{Murayama}). A ground state neutrino that has undergone a flavon interaction is assumed to be in an excited state, interaction of this excited neutrino with the Higgs boson lead to a different vacuum expectation than the original ground state neutrino which is not proportional to the absorbed energy. } 
\label{bergevinFigure2}
\end{figure*} 
 The three flavor extension then has the form{\color{black},
\begin{eqnarray}
H^{\alpha\beta\gamma}_{FD} \Psi^{\alpha\beta\gamma} = \left(\begin{matrix}
    E_{\nu_\alpha}\eta & 0 &  0   \\
    0 & E_{\nu_\beta}\eta &  0   \\
    0 & 0 & E_{\nu_\gamma}\eta   \\
  \end{matrix}\right)\Psi^{\alpha\beta\gamma},
  \label{Hamiltonian}
\end{eqnarray} 
where $\{\alpha,\beta,\gamma\}$ are the lepton flavors. This represents a system that conserves flavor for a particle of physical momentum $p$. {\color{black} This is considered the unperturbed Hamiltonian ($H^{\alpha\beta\gamma}_{FD} \equiv H_0$) and the neutrino is considered a standard Dirac neutrino particle.

 \section*{Time evolution of flavor states}
 
The  time evolution between quantum states can be derived from the unperturbed Hamiltonian (Eqn.\ref{Hamiltonian}) and the perturbation of the Flavon field (Eqn.~\ref{currents})},}
\begin{equation}
H = H_{0} + H_{1}
\end{equation}
where,
\begin{eqnarray}
    \langle\nu_{l}|H_{0}|\nu_{m}\rangle &=& \delta_{lm} E_{\nu_m} \\
    \langle\nu_{l}|H_{1}|\nu_{m}\rangle &=& (1-\delta_{lm}) {\color{black}{f_{lm}}} 
\end{eqnarray}
and $l$ or $m$ are flavor numbers ($e$, $\mu$, $\tau$), $E_{\nu_m}$ is the energy of the neutrino of flavor $m$ (Eqn.~\ref{momentumRelation}), and $f_{lm}$ is the energy width of the flavon violating interaction, $\delta_{lm}$ is the Kronecker delta. This set of interactions are illustrated in Fig.~\ref{bergevinFigure2}.}

The flavor{\color{black}-violating} interaction is not assumed to have a time dependent component ($ f_{e\mu}\delta(\vec{x})$). The system energy of an electron-neutrino and a muon-neutrino, assuming that $f_{e\mu}$ is small enough such that the system can be treated in a perturbative way and in the relativistic limit {\color{black} in the Foldy-Wouthuysen  representation }, is, 
\begin{equation}
    \begin{bmatrix}
    p+\frac{m^{2}_{\nu_{e}}}{2p} & f_{e\mu} \\
     f_{e\mu }& p+\frac{m^{2}_{\nu_{\mu}}}{2p}  
  \end{bmatrix}
  \begin{bmatrix} \nu_{e} \\ \nu_\mu
  \end{bmatrix}  \equiv  E_{sys}\begin{bmatrix} \nu_{e} \\ \nu_\mu
  \end{bmatrix}. 
  \label{basis}
 \end{equation}
where the perturbation is at least 18 orders of magnitudes smaller than the particle energy. This matrix is rewritten in the form

\begin{equation}  
E_{sys} = f_{e\mu} \sigma_x +\Delta E \sigma_z+ {E}_{tot}I, 
\end{equation}
where $E_{tot}= p+(m^{2}_{\nu_{e}}/4p)+(m^{2}_{\nu_{\mu}}/4p)$, $\Delta E = (m^{2}_{\nu_{e}}/4p) - (m^{2}_{\nu_{\mu}}/4p) $, $\sigma_x$ and $\sigma_z$ are the Pauli matrices, and $I$ is the identity matrix.  Here, $E_{sys}$ has the following eigenvalues,\footnote{ $\delta{} \equiv \sqrt{\Delta E^2+ f_{e\mu}^2}$ defined for reading simplicity}
\begin{equation}
    E_{\pm}=p+\frac{m_{\nu_e}^2}{4p}+\frac{m_{\nu_\mu}^2}{4p}\pm\delta.
\end{equation}
The relativistic Schr\"odinger equation: 
\begin{equation}
 \frac{\partial}{\partial t} \Psi = \left(-\frac{i}{\hbar} E_{sys}\right) \Psi 
 \end{equation}
 where $\Psi$ is the Weyl spinor and has solution $\Psi(t)=U(t)\Psi(0)$, where\footnote{Given two matrices, $A$ and $B$ with $[A,[A,B]] = 0$  and $[B,[A,B]]=0$:
$e^{A+B} = e^Ae^Be^{-\frac{1}{2}[A,B]}$, with $[B,A]=0$ for $A=f_{e\mu}\sigma_x+\Delta E \sigma_z$, and $B=E_{tot}I$\cite{hecht2000quantum}.},
 \begin{equation}
      U(t)= e^{-i({\color{black}{f_{e\mu}}} \sigma_x +\Delta E \sigma_z)t/\hbar}e^{-i({E}_{tot}I)t/\hbar }=U_{osc}U_{E_{\nu}},
\end{equation}
and the oscillation term can be rewritten in matrix form in the $\nu_{e}$, $\nu_{\mu}$ basis of Eqn. \ref{basis} as,

\begin{eqnarray}
U_{osc} &=&
\begin{bmatrix}
    \cos (\delta{}t) - i\frac{\Delta E}{\delta} \sin(\delta{}t) &-i\frac{f_{e\mu}}{\delta} \sin(\delta{}t)   \\
   -i\frac{f_{e\mu}}{\delta} \sin(\delta{}t)  &\cos (\delta{}t) + i\frac{\Delta E}{\delta} \sin(\delta{}t)
  \end{bmatrix}\nonumber\\
  &\equiv&
  \begin{bmatrix}
    \tilde{c}_{e\mu} & \tilde{s}_{e\mu}  \\
   \tilde{s}_{e\mu} & \tilde{\bar{c}}_{e\mu}
  \end{bmatrix}
  \label{oscillationDef}
\end{eqnarray}

\begin{figure*}[htp]
\centering
\includegraphics[width=0.98\textwidth]{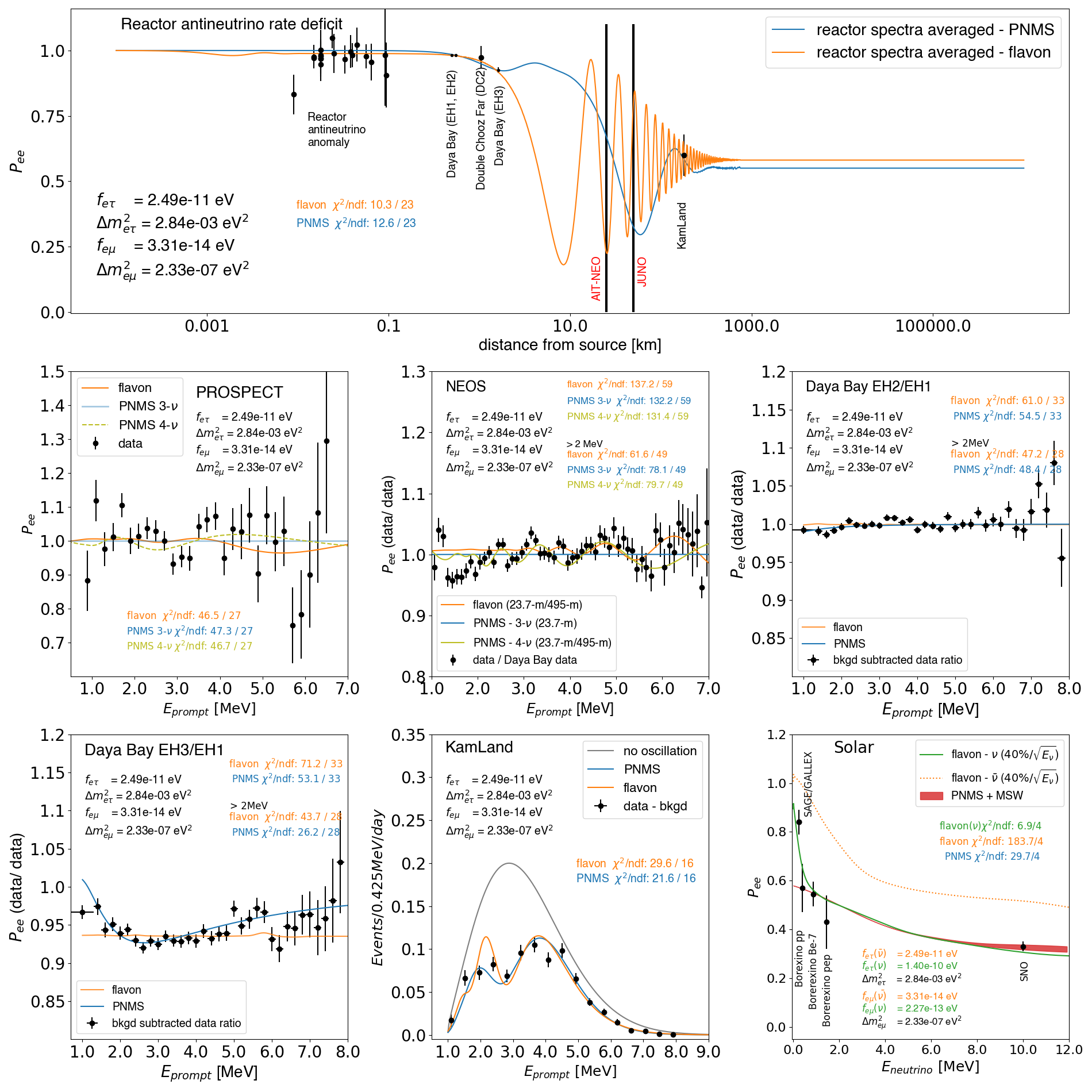} 
\hspace{2mm}
\caption{Performance of the flavon and PNMS model against a selection of electron neutrino and antineutrino experimental data. } 
\label{bergevinFigures}
\end{figure*}  

The probability that the neutrino has not changed flavor state after a time $t$ becomes,
\begin{eqnarray}
 P_{ee}\equiv P\left(\nu_e \not\rightarrow \nu_\mu\right)&=&\left|\begin{bmatrix}1\;\; 0 \end{bmatrix}U(t)  \begin{bmatrix}
  1 
  \\
  0
  \end{bmatrix}\right|^2 
\end{eqnarray}
The final form of the neutrino oscillation probability is therefore\footnote{As the mass of the neutrino is very small, momentum and the total neutrino energy are used interchangeably in this letter.},
\begin{equation}
    P_{ee} = 1
    -
    \frac
    {{\color{black}{f_{e \mu}}}^{2}}
    {\left(
    \frac{\Delta m_{\nu_{\mu e}}^{2}}{4E_\nu}\right)^{2}+{\color{black}{f_{e \mu}}}^{2}}
    \sin^{2}
    \left(t\sqrt{
    \left(
    \frac{
    \Delta m_{\nu_{\mu e}}^{2}}{4E_\nu}
    \right)^2
    +
    {\color{black}{f_{e \mu}}}^{2}}
    \right).
    \label{flavonFormula}
\end{equation}
This formula should be compared to the standard two-neutrino PNMS oscillation formula, which has the form,
\begin{equation}
    P_{ee} = 1-\sin^{2}(2\theta_{12})\sin^{2}\left(t\frac{\Delta m_{12}^{2}}{4E_{\nu}}\right).
    \label{PNMSFormula}
\end{equation}
Each formulas has the same number of physical constants and as such the flavon model is not more complex than the PNMS model. If one wishes, an energy-dependent mixing angle term for the flavon formula could be define as: 
\begin{equation}
    \theta_{e\mu}'(E_\nu;\Delta m_{e\mu}^2,f_{e\mu}) = \frac{1}{2}\sin^{-1}\left( \frac{f_{e\mu}^2}{\left(\frac{\Delta m_{e\mu}^2}{4E_\nu}\right)^2+f_{e\mu}^2} \right),
    \label{theta}
\end{equation}
leading Eqn. \ref{flavonFormula} to be re-written as,
\begin{equation}
    P_{ee} = 1
    -
 \sin^2(2\theta'_{e\mu})\sin^{2}
    \left(t\sqrt{
    \left(
    \frac{
    \Delta m_{\nu_{\mu e}}^{2}}{4E_\nu}
    \right)^2
    +
    f_{e \mu}^{2}}
    \right).
    \label{flavonFormula2}
\end{equation}
For experiments at very short or very long ranges, the neutrino gains non-standard oscillation properties. At the very short range,  Eqn. \ref{flavonFormula} simplifies to ($1-f_{e\mu}^2t^2$). While at very long range and at sufficient energy, Eqn. \ref{flavonFormula} simplifies to $0.5\left(1+(\Delta m_{\alpha\beta}^2/4E_{\nu}f_{\alpha\beta})^2\right)$, leading to low-energy excess that do not follow the expected PNMS oscillation predictions. 
\section*{Testing the model}
A global fit using a simple $\Delta\chi^{2}$ method between this model and published experimental data is made in order to study neutrino and antineutrino data across multiple experimental conditions. Generally, oscillations are observed using sources of (anti)neutrinos of a specific flavor (i.e., reactor electron antineutrinos, solar electron neutrinos, beam muon neutrinos) and measuring how neutrinos transform (or do not transform) into other flavors as a function of baseline and energy. This observation is made by either measuring how many neutrinos are lost---in the case of a disappearance experiment---or how many neutrinos of a different flavor are created---in the case of an appearance experiment. 
It is assumed as a first order approximation that the extension to the three neutrino survival probability for an electron neutrino is simply{\color{black}:
\begin{eqnarray}
    P_{ee} &=& |\langle e | R_x(\delta_{\mu\tau}'t)R_y(\delta_{e\tau}'t)R_z(\delta_{e\mu}'t) |e\rangle |^2\\
    P_{ee} &\approx& |\langle e|\tilde{c}_{e\mu}\tilde{c}_{e\tau}|e\rangle|^2= P^{2D}_{\nu_e\not\rightarrow \nu_\mu} P^{2D}_{\nu_e\not\rightarrow\nu_\tau}\label{pmue}
\end{eqnarray}
where $R_i(\delta'_{ij}t)$ is the rotation along the flavor axis $i$, and $\tilde{c}_{e\mu}$ are from Eq.~\ref{oscillationDef}. We note that $R_x R_y R_z \neq R_y R_x R_z$ and as such the oscillation formula is only approximately correct for the disappearance experiments relevant to this paper.}

The top row of Fig.~\ref{bergevinFigures} shows the rate of reactor antineutrino disappearance\footnote{averaged over reactor energy spectra} as a function of distance from creation\footnote{Values above 750 km are averaged over remaining baseline due to numerical computation issues.}. The central values for the antineutrino parameters are displayed in this and each other sub figure for the flavon model. The short-baseline and Double Chooz data \cite{PhysRevD.83.073006,PhysRevLett.108.131801}, the Daya Bay \cite{PhysRevD.95.072006} results were adjusted for recent flux results including a 3\% upward shift.   We note that the change in average disappearance value between the Daya Bay near detectors and far detector is consistent with the oscillation parameters of KamLand and is the result of the non-standard oscillation behavior of the flavon model at short baseline. In the PNMS model a third neutrino is needed to explain this deficit, which in turn require a fourth neutrino to explain the short-baseline anomaly. There is no need to extend the number of neutrinos for the case of the 3-neutrino flavon model.
\begin{figure*}[htp]
\centering
\includegraphics[width=0.85\textwidth]{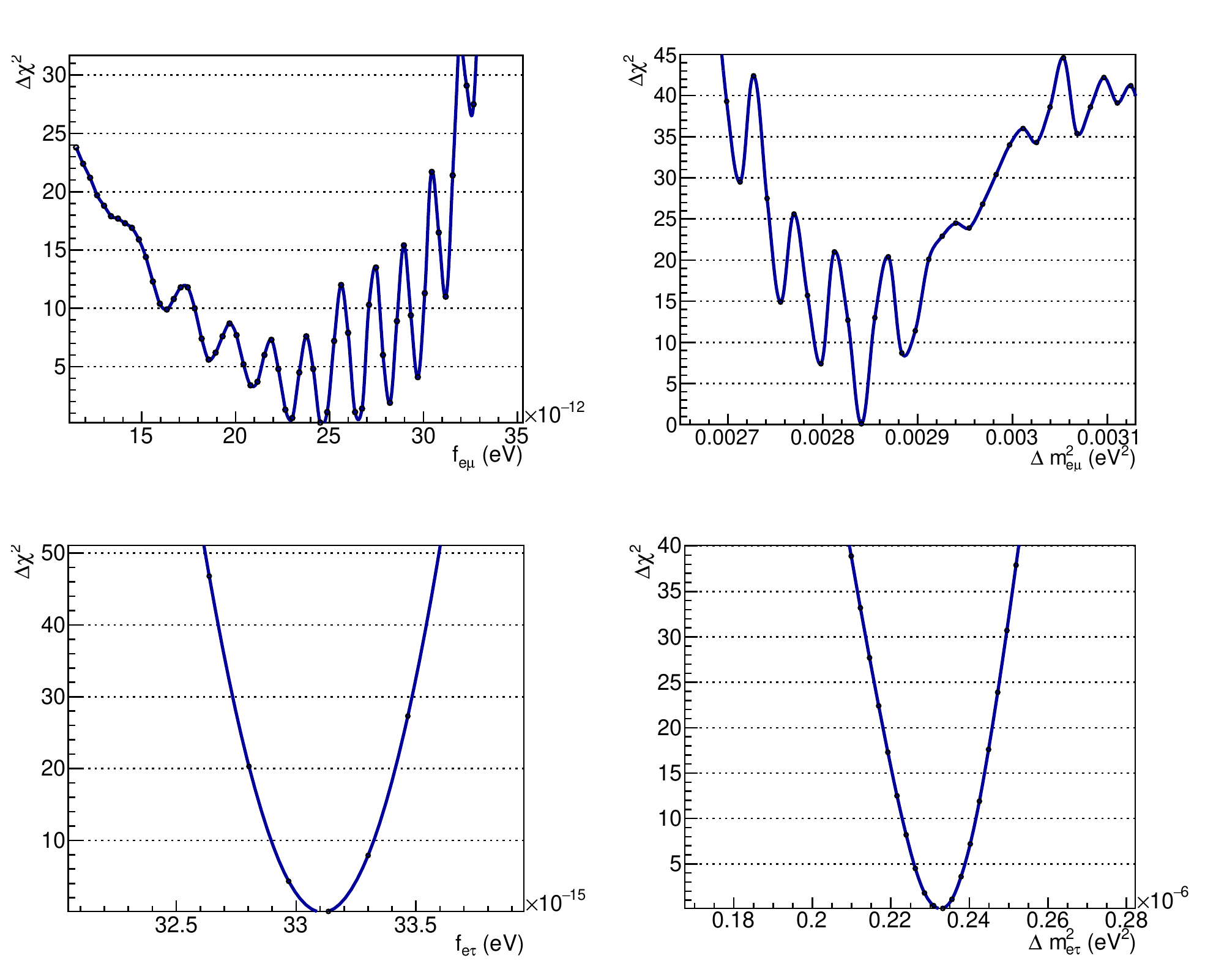} 
\caption{The $\Delta\chi^{2}$ map of the reactor antineutrino best fit from Fig.~\ref{bergevinFigures}. The $\Delta \chi^{2}$ for a parameter of interest is obtained by fixing the value of the best fit {\color{black} of} the three other nuisance parameters.} 
\label{bergevinFigure4}
\end{figure*}
The second row of Fig.~\ref{bergevinFigures} shows the spectral oscillations in the electron to tau neutrino conversion. Oscillations at PROSPECT \cite{PhysRevLett.121.251802} were evaluated using the publicly available detector response matrix. The probability prediction are further corrected by normalising to unity. Oscillation at NEOS \cite{PhysRevLett.118.121802} were evaluated using the data ratio NEOS/Daya Bay is presented including the best fit parameter for sterile neutrinos. Oscillations at Daya Bay for the two near detectors (EH1 and EH2) site are compared to data from \cite{PhysRevD.95.072006} assuming a detector resolution of 7\%/$\sqrt{E}$ and accounting for respectively thirty-six and forty-eight reactor-detector baselines over two data taking period. Data ratios were performed in order to remove uncertainties on flux and cross-section models and background were subtracted according to the best fit values prior to taking the data ratio. 

The third row of Fig.~\ref{bergevinFigures} shows the spectral oscillation in the electron to muon conversion. Oscillations at Daya Bay using the far hall and the near hall (EH3 and EH2) are again used to remove uncertainties on flux and cross sections. The spectral feature at 5.5-6.0 MeV is due to detector resolution effects. Oscillations at KamLand were compared to data from \cite{PhysRevD.88.033001}, where we assumed an energy resolution of 7\%/$\sqrt{E}$. The oscillation probability was evaluated based on published power information for five reactor sites\footnote{Kashiwazaki, Ohi, Takahama, Hamaoka, and Tsuruga.}. The $^{210}$Po, accidentals and geoneutrinos were removed in the background-subtracted data for the first phase of \cite{PhysRevD.88.033001}. The last figure shows the oscillation from solar neutrinos and how the flavon antineutrino-parameters {\color{black}(dotted black line)} do not fit the solar data well, but a set of two different flavon {\color{black} perturbations} fits the data including the SAGE-GALLEX point. {\color{black} The MSW-LSA effect as illustrated in the red band is not considered in the flavon model, nonetheless the MSW behavior is reproduced by the flavon model, as is demonstrated by the  agreement of the red band and green line. This is} due to the non-standard oscillation behavior at long baseline.
\begin{figure*}[htp]
\centering
\includegraphics[width=0.85\textwidth]{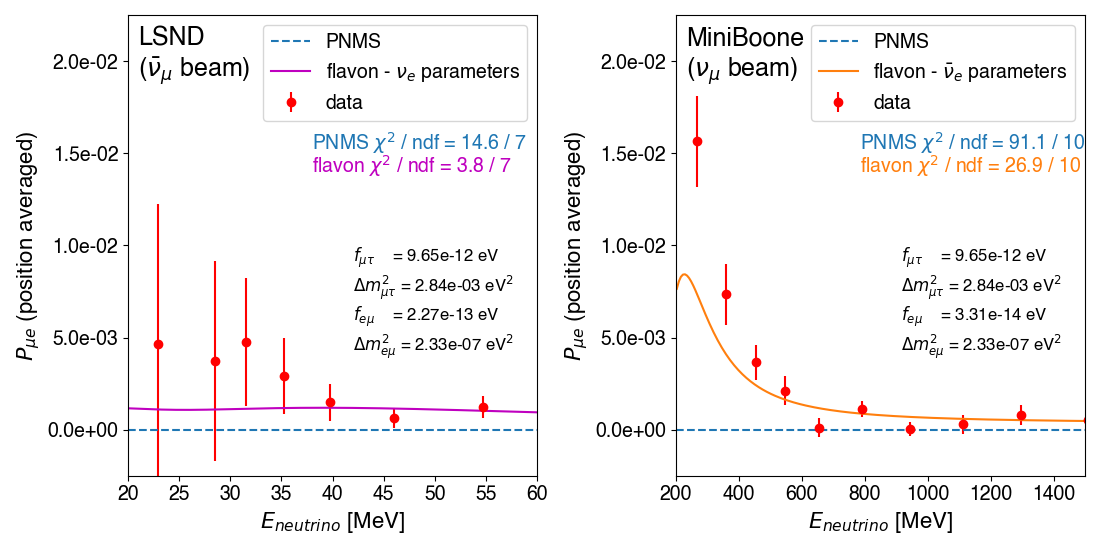} 
\caption{Preliminary {\color{black} fit} results for {\color{black} the LSND and MiniBoone experimental data with a common flavon perturbation $f_{\mu\tau}$, with other parameters fixed. }}
\label{bergevinFigure3}
\end{figure*} 

In all cases, the flavon fit results are largely consistent with the PNMS results. Disagreements can be observed in a region below 2 MeV for some experiments, such as in the NEOS experiment, KamLand, and the far Daya Bay results. It should be noted that a variety of backgrounds populate the energy region below 2 MeV, and more precise treatment should be performed in lieu of the background subtraction method used in this article. {\color{black}   Fig.~\ref{bergevinFigure4} shows the global fit results for the antineutrino experiments considered. The global best set of fit values is shown on each plot of Fig.~\ref{bergevinFigures}. 

Fig.~\ref{bergevinFigure3} shows experimental data in the beam sector that are considered anomalous in the PNMS model but fit well the flavon model for perturbation energies and mass square difference found in the solar and reactor antineutrino sectors. This data is considered anomalous as there should be no signal observed in this energy range. At LSND for the mode $\bar{\nu}_\mu\rightarrow\bar{\nu}_{e}$ good fits are observed when the perturbation from the Solar data is used. For MiniBoone, fit results are sensible for ${\nu}_\mu\rightarrow{\nu}_{e}$ when using the reactor flavor $f_{e\mu}$ perturbation. Reasons for the order of such transitions order is not yet understood. The flavon prediction with the solar $f_{e\mu}$ is not in agreement with the $\bar{\nu}_\mu\rightarrow\bar{\nu}_{e}$ expectation (not shown here) for MiniBoone.} 

We show hints that left-handed neutrino and right-handed antineutrinos oscillate differently as the antineutrino parameters cannot reproduce the solar data, but a set of different parameters can. This, in turn, implies CP violation, which is one of three of the Sakharov conditions proposed to explain why our universe is dominated by matter. Since the neutrino data considered here consists of only four experimental measurements, no official CP-violation claim can be made. 

\section*{Conclusions and Potential Impact}
 
In this letter a new oscillation model is proposed (Eqn. \ref{flavonFormula}) as a replacement to the PNMS formalism (Eqn. \ref{PNMSFormula}). This model recreates oscillation features measured in previous experiments using a simple $\Delta\chi^2$ method. However, more sophisticated fitting techniques and new oscillation data would be required to better test this model.  Of particular interest are the accelerator and atmospheric neutrino sectors where preliminary agreements are observed (Fig.\ref{bergevinFigure3}). A follow-up paper will discuss the sensitivity of future planned experiments to this model and explore detector observable effects in beam-line experiments. It should be already noted from Fig.~\ref{bergevinFigures} (top) that the AIT{\color{black}-NEO} and JUNO experiments will be sensitive to this model as they will be respectively at an oscillation minimum and maximum and will further constrain and confirm $f_{e\mu}$ and $\Delta m^{2}_{e\mu}$, current results as shown in Figure \ref{bergevinFigure4}.

Further studies of the possible implications to the non-proliferation neutrino community are being investigated. As the oscillation patterns are more complex than in the standard PNMS model, one can potentially imagine leveraging the observed neutrino spectra for multiple purposes---either to make more confident pronouncements as to whether a reactor complex is complying with declared operations or to verify compliance with future treaties to verify the absence of undeclared reactors via observed changes in the energy spectra. 

Neutrinos and antineutrinos are found to oscillate with different strengths in the flavon model between the solar and reactor sector, this implies charge-parity violation for neutrinos. This is a key requirement of the Sakharov conditions, which are necessary for understanding the matter-antimatter imbalance observed in our universe. 

For some time, the physics community has {\color{black} operated under the premise that charged and neutral leptons have different underlying properties. Charged leptons are produced in weak interactions as single mass states, and neutral leptons are produced as superpositions of mass states}. At the price of assuming a simple interaction with a dark matter field, and a straightforward re-casting of the equations governing neutrino oscillations to include an energy-dependent term, this letter presents a model for oscillations with a single valued mass for each neutrino flavor{\color{black}, consistent with the charged lepton picture}. This model agrees with a diverse set of experimental results, as well as, or better than, the prevailing PNMS model. 
\begin{acknowledgements}
 The author would like to thank Bernie Nickel for his advice when the author was developing a similar oscillation model for the neutron. The author would like to thank Baha Balantekin, Adam Bernstein, Nathaniel Bowden, Steven Dazeley, Ferenc Dalnoki-Veress, Christopher Grant, Viacheslav Li, and Michael Mendenhall for discussions related to certain aspects of this work. The author is appreciative to the DAYA BAY and PROSPECT collaborations for publishing their data online. This work was performed under the auspices of the U.S. Department of Energy by Lawrence Livermore National Laboratory under Contract DE-AC52-07NA27344, LLNL-JRNL-772746.
 
 \end{acknowledgements}

\bibliographystyle{spphys}

\end{document}